\documentclass[aip,jcp,preprint,groupedaddress,superscriptaddress]{revtex4-2}
\usepackage{graphicx}
\usepackage{color}
\usepackage{xcolor}
\usepackage{amsmath}
\usepackage{amsfonts}
\usepackage{amssymb}
\usepackage{float}

\begin{document}
\title{Equilibrium-nonequilibrium ring-polymer molecular dynamics for nonlinear spectroscopy}
\author{Tomislav Begu\v{s}i\'{c}}
\email{tbegusic@caltech.edu}
\author{Xuecheng Tao}
\affiliation{Division of Chemistry and Chemical Engineering, California Institute of Technology, Pasadena, California 91125, USA}
\author{Geoffrey A. Blake}
\affiliation{Division of Chemistry and Chemical Engineering, California Institute of Technology, Pasadena, California 91125, USA}
\affiliation{Division of  Geological and Planetary Sciences, California Institute of Technology, Pasadena, California 91125, USA}
\author{Thomas F. Miller III}
\affiliation{Division of Chemistry and Chemical Engineering, California Institute of Technology, Pasadena, California 91125, USA}
\date{\today}

\begin{abstract}
Two-dimensional Raman and hybrid terahertz-Raman spectroscopic techniques provide invaluable insight into molecular structure and dynamics of condensed-phase systems. However, corroborating experimental results with theory is difficult due to the high computational cost of incorporating quantum-mechanical effects in the simulations. Here, we present the equilibrium-nonequilibrium ring-polymer molecular dynamics (RPMD), a practical computational method that can account for nuclear quantum effects on the two-time response function of nonlinear optical spectroscopy. Unlike a recently developed approach based on the double Kubo transformed (DKT) correlation function, our method is exact in the classical limit, where it reduces to the established equilibrium-nonequilibrium classical molecular dynamics method. Using benchmark model calculations, we demonstrate the advantages of the equilibrium-nonequilibrium RPMD over classical and DKT-based approaches. Importantly, its derivation, which is based on the nonequilibrium RPMD, obviates the need for identifying an appropriate Kubo transformed correlation function and paves the way for applying real-time path-integral techniques to multidimensional spectroscopy.
\end{abstract}

\maketitle

\section{Introduction}

Two-dimensional vibrational spectroscopy is a versatile technique to study microscopic interactions at their natural, femtosecond time scales.\cite{Hamm_Zanni:2011} Recently, a series of hybrid spectroscopic experiments\cite{Hamm_Savolainen:2012, Savolainen_Hamm:2013,Finneran_Blake:2016,Grechko_Bonn:2018} involving mid-infrared, far-infrared (or terahertz), and visible (Raman) pulses have been developed to study electrical and mechanical anharmonicities,\cite{Hamm:2014,Finneran_Blake:2017,Magdau_Miller:2019,Mead_Blake:2020} structural heterogeneities of liquids,\cite{Shalit_Hamm:2017,Hamm_Shalit:2017a,Berger_Shalit:2019} and the couplings between intermolecular and intramolecular vibrational modes.\cite{Ciardi_Shalit:2019,Vietze_Grechko:2021} However, the interpretation of such spectra is still an open question and requires adequate simulation methods.\cite{Hamm:2019,Berger_Shalit:2019,Sidler_Hamm:2019,Sidler_Hamm:2020}

Several computational methods have been proposed for simulating two-dimensional off-resonant Raman and hybrid terhertz--Raman spectra. Model-based approaches aim at constructing simplified, few-dimensional model systems that can be solved fully quantum mechanically in the presence of a harmonic oscillator bath.\cite{Tanimura_Ishizaki:2009,Ikeda_Tanimura:2015} These approaches can disentangle contributions to spectra and relate them to different physical effects encoded in the model parameters. The parameters can be obtained by fitting to the existing experiment\cite{Magdau_Miller:2019} or molecular dynamics (MD) simulations.\cite{Ueno_Tanimura:2020} Alternatively, MD can be used to simulate spectra directly,\cite{Jansen_Duppen:2000,Jansen_Duppen:2001,Saito_Ohmine:2002,Saito_Ohmine:2003,DeVane_Keyes:2003,Hasegawa_Tanimura:2006,Hasegawa_Tanimura:2008,Yagasaki_Saito:2009,Ito_Tanimura:2014,Hamm:2014,Sun:2019,Jansen_Cho:2019} as a way to cross-validate the simplified model and further test the molecular mechanics force field or ab initio quantum chemistry method used for describing the forces between atoms, dipole moments, and polarizabilities.\cite{Hasegawa_Tanimura:2011,Ito_Tanimura:2016,Sidler_Hamm:2018} However, one improvement of MD is highly desired---MD simulations are based on classical atomic nuclei and neglect nuclear quantum effects, which can significantly modify the spectra. Indeed, recent two-dimensional Raman--THz--THz spectra of H$_2$O and D$_2$O revealed the effect of isotopic substitution beyond what would be expected from classical mechanics, pointing at nuclear quantum effects of the light hydrogen atoms.\cite{Berger_Shalit:2019}

Quantum simulations of the condensed phase have been enabled by various semiclassical methods based on the classical MD framework, including linearized semiclassical initial value representation (LSC-IVR),\cite{Wang_Miller:1998,Liu_Miller:2006} path-integral Liouville dynamics,\cite{Liu:2014} centroid molecular dynamics (CMD),\cite{Jang_Voth:1999} and ring-polymer molecular dynamics (RPMD).\cite{Craig_Manolopoulos:2004} These methods have proven useful for the computation of one-time correlation functions\cite{Habershon_Miller:2013} related to various dynamical properties, including reaction rates,\cite{Wang_Miller:1998,Jang_Voth:2000,Boekelheide_Miller:2011,Suleimanov_Guo:2016,Tao_Miller:2019,Tao_Miller:2020} diffusion coefficients,\cite{Miller_Manolopoulos:2005,Miller_Manolopoulos:2005a} and one-dimensional vibrational spectra.\cite{Witt_Marx:2009,Rossi_Manolopoulos:2014,Benson_Althorpe:2019,Trenins_Althorpe:2019,Korol_Miller:2020,Rosa-Raices_Miller:2021} Some of them have also been applied to two-dimensional electronic\cite{Loring:2017,Polley_Loring:2020,Provazza_Coker:2018,Provazza_Coker:2020,Gao_Geva:2020} and infrared spectroscopies,\cite{Gruenbaum_Loring:2009,Gerace_Loring:2013,Alemi_Loring:2015,Kwac_Geva:2013} in which the quantum subsystem, consisting of, e.g., electronic or high-frequency vibrational degrees of freedom, can be well defined.\cite{Shi_Geva:2008,McRobbie_Geva:2009a,McRobbie_Geva:2009b} In contrast, their application to two-dimensional Raman and hybrid terahertz-Raman spectroscopic techniques, in which all vibrational degrees of freedom should be treated on an equal footing, has been limited. Recently, the group of Batista\cite{Jung_Batista:2018,Jung_Batista:2019,Jung_Batista:2020} proposed a set of methods that approximate the symmetric contribution to the so-called double Kubo transformed (DKT) correlation function, which is a two-time extension of the original one-time Kubo transformed correlation function.\cite{Habershon_Miller:2013} However, the relation between the symmetrized DKT correlation function and the spectroscopic response function is only approximate.

Here, we present an alternative RPMD approach to two-dimensional spectroscopy, which directly considers the two-time response function and does not rely on the additional approximation related to the DKT correlation function. To derive the proposed method, we start from the recently developed nonequilibrium RPMD\cite{Welsch_Miller:2016,Marjollet_Welsch:2020,Marjollet_Welsch:2021,Jiang_Bunermann:2021,Marjollet_Welsch:2022} and employ classical response theory. We then explore its validity and limitations both theoretically and numerically.

\section{Theory}

In two-dimensional Raman\cite{Tanimura_Mukamel:1993,Palese_Miller:1994} and hybrid terahertz-Raman spectroscopies,\cite{Savolainen_Hamm:2013,Finneran_Blake:2016,Grechko_Bonn:2018} the signal is measured as a function of two delay times between three ultrashort light pulses, whose specific sequence determines the type of spectroscopy. To keep the discussion general, we consider the time-dependent Hamiltonian
\begin{equation}
\hat{H}_{\text{tot}} = \hat{H} - \hat{A} F_1(t) - \hat{B} F_2(t) \label{eq: H_tot}
\end{equation}
comprised of the system's field-free Hamiltonian $\hat{H}$ and two interactions with the first two ultrashort pulses, controlled, respectively, by coordinate-dependent operators $\hat{A}$ and $\hat{B}$ and by time-dependent functions $F_{1,2}(t)$ representing the pulse shapes. For terahertz pulses, the interaction operators are the system's dipole moments and functions $F(t)$ are the electric fields of the light, whereas for the visible/near-infrared (Raman) pulses or sum-frequency terahertz excitation,\cite{Mead_Blake:2020} the operators are the system's polarizabilities and the time-dependent functions correspond to the squares of the electric fields. $F_1$ is centered at $t = -t_1$, $F_2$ is centered at $t=0$, and the signal is measured at $t = t_2$, where $t_1$ and $t_2$ represent the time delays between the three light pulses. The recorded signal is proportional to the expectation value\cite{Finneran_Blake:2017}
\begin{equation}
\langle \hat{C}(t_2) \rangle = \text{Tr}[\hat{C} \hat{\rho}(t_2)] \label{eq:C_exp}
\end{equation} 
of another coordinate-dependent operator $\hat{C}$, which is again either the polarizability or dipole moment operator of the system. In Eq.~(\ref{eq:C_exp}), system's equilibrium density operator $\hat{\rho} = \exp(-\beta \hat{H}) / \text{Tr}[\exp(-\beta \hat{H})]$ evolves under the time-dependent Hamiltonian (\ref{eq: H_tot}), i.e.,
\begin{eqnarray}
\hat{\rho}(t) &=& \hat{U}_{\text{tot}}(t, -\infty) \hat{\rho} \hat{U}_{\text{tot}}^{\dagger}(t, -\infty), \label{eq:rho_t_tot}\\
\hat{U}_{\text{tot}}(t_f, t_i) &=& \mathcal{T} \exp \left[ -\frac{i}{\hbar} \int_{t_i}^{t_f} d\tau \hat{H}_{\text{tot}}(\tau) \right],\label{eq:U_tot}
\end{eqnarray}
$\beta$ is the inverse temperature, and $\mathcal{T}$ is the time ordering operator. In practice, the experiments are designed to extract the components of the signal that are proportional to $F_1$ and $F_2$, which can be accounted for by evaluating the signal as\cite{Magdau_Miller:2019}
\begin{equation}
S(t_2) = \langle \hat{C}(t_2) \rangle_{++} - \langle \hat{C}(t_2) \rangle_{+-} - \langle \hat{C}(t_2) \rangle_{-+} + \langle \hat{C}(t_2) \rangle_{--}, \label{eq:S_nonpert}
\end{equation}
where $\langle \cdot \rangle_{jk}$ represents the expectation value (\ref{eq:C_exp}) of the system evolved under the Hamiltonian (\ref{eq: H_tot}) with $F_1(t) \rightarrow (j/2) F_1(t)$ and $F_2(t) \rightarrow (k/2) F_2(t)$. For sufficiently weak external fields, we can further invoke second-order time-dependent perturbation theory and recover the well-known result\cite{Tanimura_Mukamel:1993}
\begin{equation}
S(t_2) = \int_{0}^{t_2} d\tau_2 \int_{0}^{\tau_2} d\tau_1 R(\tau_2, \tau_1) F_2(t_2-\tau_2) F_1(t_2-\tau_2 - \tau_1), \label{eq:S_pert}
\end{equation}
where
\begin{equation}
R(\tau_2, \tau_1) = -\frac{1}{\hbar^2} \text{Tr}\left\lbrace \hat{C}(\tau_2 + \tau_1) [\hat{B}(\tau_1), [\hat{A}, \hat{\rho}]] \right\rbrace \label{eq:R_qm_pert}
\end{equation}
is the corresponding response function, which depends only on the system's properties. Because it can be easily convolved with different experimental pulses to recover the observed experimental signals (\ref{eq:S_pert}), the response function is often the direct target of many theoretical studies. Here, we also intend to focus on the response function, but take a slightly different route in developing the computational approach. Namely, we note that for delta pulses, $F_1(t) = \varepsilon_1 \delta(t + t_1)$ and $F_2(t) = \varepsilon_2  \delta(t)$, the signal measured after the second time delay $t_2$ is proportional to the response function,
\begin{equation}
S(t_2) = \varepsilon_1 \varepsilon_2 R(t_2, t_1), \label{eq:S_pert_delta}
\end{equation}
where $\varepsilon_{1,2}$ control the amplitude of the external fields. In the following, we derive an approximation to $S(t_2)$ under weak delta pulses and relate it to $R(t_2, t_1)$ using Eq.~(\ref{eq:S_pert_delta}).

To develop a tractable theory for condensed-phase simulations of spectra, we replace the exact quantum-mechanical expression (\ref{eq:C_exp}) by its nonequilibrium RPMD\cite{Welsch_Miller:2016} approximation
\begin{equation}
\langle \hat{C}(t) \rangle_{jk} \approx \langle C(t) \rangle^{\text{RP}}_{jk} = \int dq \int dp \ C_{N}(q_{jk,t}) \rho_N(q,p), \label{eq:C_RP_NEQ}
\end{equation}
where $q$ and $p$ are the positions and momenta of the extended system comprised of $N$ replicas of the original $D$-dimensional system,
\begin{equation}
\rho_N(q,p) = \frac{e^{-\beta_N H_N (q, p)}}{\int dq \int dp\ e^{-\beta_N H_N (q, p)}}
\end{equation}
is the corresponding phase-space distribution,
\begin{equation}
H_N(q, p) = \sum_{i=1}^{N} \frac{1}{2} p_i^T \cdot m^{-1} \cdot p_i + \frac{1}{2(\beta_N \hbar)^2}(q_i - q_{i-1})^T \cdot m \cdot (q_i - q_{i-1}) + V(q_i) \label{eq:H_N}
\end{equation}
is the system's field-free ring-polymer Hamiltonian,\cite{Habershon_Miller:2013} $\beta_N = \beta / N$,
\begin{equation}
C_N(q) = \frac{1}{N} \sum_{i=1}^{N} C(q_i),
\end{equation}
$m$ is the symmetric mass matrix of the system, and $q_0 = q_N$. The classical time evolution of $q_{jk,t}$ and $p_{jk,t}$ is governed by the time-dependent ring-polymer Hamiltonian
\begin{eqnarray}
H_{jk,N}(q, p, t) = H_N(q, p) + j V_{A,N}(q) \delta(t+t_1) + k V_{B,N}(q) \delta(t),\label{eq:H_jk_N}\\
V_{A,N}(q) = -\frac{\varepsilon_1 }{2} N A_N(q),\quad V_{B,N}(q) = -\frac{\varepsilon_2}{2}N B_N(q), 
\end{eqnarray}
with $j, k \in \lbrace +, - \rbrace$. Nonequilibrium RPMD, in which the dynamics and initial distribution depend on different Hamiltonians [see Eqs.~(\ref{eq:C_RP_NEQ})--(\ref{eq:H_N}) and (\ref{eq:H_jk_N})], has been rigorously justified as an approximation to the more general real-time nonequilibrium Matsubara dynamics.\cite{Welsch_Miller:2016} Although its original derivation involved only time-independent Hamiltonians, this assumption was not used explicitly at any step of the derivation, justifying the use of nonequilibrium RPMD even in the time-dependent setting. At this point, we recognize that the equations above already formulate a valid computational technique for evaluating the response function, which, in the limit of $N=1$, is equivalent to the finite-field nonequilibrium MD method of Jansen, Snijders, and Duppen.\cite{Jansen_Duppen:2000,Jansen_Duppen:2001} In contrast to their classical approach, the nonequilibrium RPMD theory can include, at least approximately, the nuclear quantum effects on two-dimensional spectra when the path-integral continuum limit is reached. However, these nonequilibrium (RP)MD methods are impractical because a different set of trajectories is needed for each choice of the delay time $t_1$. To address this problem, we employ the equilibrium-nonequilibrium approach, originally developed by Hasegawa and Tanimura\cite{Hasegawa_Tanimura:2006} in the context of classical MD simulations, in which one of the two pulses is treated perturbatively, and combine it with the RPMD to account for nuclear quantum effects.

We start by rewriting Eq.~(\ref{eq:C_RP_NEQ}) as
\begin{equation}
\langle C(t) \rangle^{\text{RP}}_{jk} = \int dq \int dp\ C_{N}(q) U_{jk,N}(t, -\infty) \rho_N(q,p),
\end{equation}
where
\begin{equation}
U_{jk, N}(t_f, t_i) = \mathcal{T} \exp\left[-\int_{t_i}^{t_f} d\tau L_{jk, N}(\tau)\right]
\end{equation}
governs the time evolution under the Hamiltonian $H_{jk, N}$, 
\begin{equation}
L_{jk, N}(t) = \lbrace \cdot , H_{jk, N}(t)\rbrace = L_N + j \delta(t + t_1) L_{A,N} + k \delta(t) L_{B,N}
\end{equation}
is the corresponding Liouvillian, $L_N =  \lbrace \cdot , H_{N}\rbrace$, $L_{A, N} =  \lbrace \cdot , V_{A, N}\rbrace$, $L_{B, N} =  \lbrace \cdot , V_{B, N}\rbrace$, and $\lbrace \cdot, \cdot \rbrace$ denotes the Poisson bracket. Next, we expand\cite{Holian_Evans:1985,Ple_Bonella:2021}
\begin{eqnarray}
U_{jk,N}(t, -\infty) 
&\approx& U_{k,N}(t, -\infty) - j \int_{-\infty}^{t} d\tau U_{k, N}(t,\tau) L_{A,N} \delta(\tau+t_1) U_{k,N}(\tau, -\infty) \label{eq:U_jk_LR_1}\\
&=& U_{k,N}(t, -\infty) - j U_{k, N}(t,-t_1) L_{A,N} U_{N}(-t_1, -\infty) \label{eq:U_jk_LR_2}
\end{eqnarray}
to the first order in $\varepsilon_1$ to obtain
\begin{equation}
\langle C(t) \rangle^{\text{RP}}_{jk} \approx \langle C(t) \rangle^{\text{RP}, (0)}_{k} + \langle C(t) \rangle^{\text{RP},(1)}_{jk}, \label{eq:C_jk_approx}
\end{equation}
where
\begin{equation}
\langle C(t) \rangle^{\text{RP},(0)}_{k} = \int dq \int dp\ C_{N}(q) U_{k,N}(t, -\infty) \rho_N(q,p) \label{eq:C_0_jk}
\end{equation}
and
\begin{equation}
\langle C(t) \rangle^{\text{RP},(1)}_{jk} 
= -j \int dq \int dp\ C_{N}(q) U_{k,N}(t, -t_1) L_{A,N} U_{N}(-t_1, -\infty) \rho_N(q,p).
\label{eq:C_1_jk}
\end{equation}
In Eq.~(\ref{eq:U_jk_LR_1}) we introduced a short-hand notation
\begin{equation}
U_{k, N}(t_f, t_i) = \mathcal{T} \exp\left\lbrace-\int_{t_i}^{t_f} d\tau [L_{N} + k \delta(\tau)L_{B,N}]\right\rbrace
\end{equation}
for the evolution operator that involves only the second pulse. In going from Eq.~(\ref{eq:U_jk_LR_1}) to Eq.~(\ref{eq:U_jk_LR_2}), we assumed that $t, t_1 > 0$ and used
\begin{equation}
U_{k,N}(t_f, t_i) = 
\begin{cases} 
      U_N(t_f, t_i), & 0 \notin (t_i, t_f), \\
      U_{k,N}(t_f,0) U_N(0, t_i), & \text{otherwise},
\end{cases}
\label{eq:U_k_cases}
\end{equation}
with
\begin{equation}
U_N(t_f,t_i) = e^{-L_N(t_f-t_i)}.
\end{equation}
Then, we derive
\begin{equation}
\langle C(t) \rangle^{\text{RP},(1)}_{jk} =  \frac{j \beta \varepsilon_1}{2} \int dq \int dp\  C_{N}(q_{k, t}) \dot{A}_N(q_{-t_1}) \rho_N(q,p)
\label{eq:C_1_jk_final}
\end{equation}
from Eq.~(\ref{eq:C_1_jk}) by applying $U_{N}(-t_1, -\infty) \rho_N(q,p) = \rho_N(q,p)$,
\begin{eqnarray}
L_{A,N} \rho_N(q,p) 
&=& - \left[\frac{\partial V_{A,N}(q)}{\partial q}\right]^{T} \cdot \frac{\partial}{\partial p} \rho_N(q,p) \\
&=& - \beta_N \frac{\varepsilon_1}{2} \sum_{i=1}^{N} \left[\frac{\partial A(q_i)}{\partial q_i}\right]^{T} \cdot \dot{q}_i \rho_N(q,p) \\
&=& - \beta \frac{\varepsilon_1}{2} \dot{A}_N(q) \rho_N(q,p),
\end{eqnarray}
where $\dot{f} = \partial f / \partial t$ denotes the time derivative, and Eq.~(\ref{eq:U_k_cases}) in combination with $C_N(q_{k,t}) = C_N(q) U_{k,N}(t,0)$ and $\dot{A}_N(q_{-t_1}) = U_N(0,-t_1) \dot{A}_N(q)$. $q_{-t_1}$ is the position along a backward equilibrium trajectory. $q_{k,t_2}$ corresponds to a nonequilibrium trajectory evolved under the Hamiltonian that involves the interaction with the second pulse, which is equivalent to the field-free evolution with the initial conditions $q_{k,0} = q$ and $p_{k,0} = p + (k \varepsilon_2 / 2) N [\partial B_N(q) / \partial q]$.

Combining Eq.~(\ref{eq:C_1_jk_final}) with Eqs.~(\ref{eq:S_nonpert}), (\ref{eq:S_pert_delta}), and (\ref{eq:C_jk_approx}) yields the final result
\begin{equation}
R(t_2, t_1) \approx \frac{\beta}{\varepsilon_2} \int dq \int dp [ C_N(q_{+, t_2}) - C_N(q_{-, t_2})] \dot{A}_N(q_{-t_1}) \rho_N(q,p), \label{eq:R_RPMD}
\end{equation}
assuming $\varepsilon_2$ is sufficiently small to guarantee the validity of the perturbative expression (\ref{eq:S_pert_delta}). It follows that the response function $R(t_2, t_1)$ can be evaluated as an equilibrium ensemble average of an estimator based on two nonequilibrium trajectories ($q_{\pm, t_2}$) and one backward equilibrium trajectory ($q_{-t_1}$). At high temperatures, where $\beta \rightarrow 0$, or for $N=1$, Eq.~(\ref{eq:R_RPMD}) reduces to the equilibrium-nonequilibrium classical MD approach.\cite{Hasegawa_Tanimura:2006} In contrast to the classical approach, the RPMD method captures nuclear quantum effects, in accordance with the well-known advantages and limitations of the RPMD for one-time correlation functions.\cite{Habershon_Miller:2013} Furthermore, in the Supplementary Material, we show that the RPMD is exact for the harmonic oscillator when two out of three operators $\hat{A}$, $\hat{B}$, and $\hat{C}$ are linear functions of position. Finally, the proposed equilibrium-nonequilibrium RPMD method always satisfies $R(t_2=0, t_1) = 0$ (because $q_{+,0} = q_{-,0}$), which holds for the exact quantum response function (\ref{eq:R_qm_pert}), but is not guaranteed by some other approximate methods (as discussed below).

Before demonstrating these properties numerically, we briefly review the only other RPMD-based method proposed to date for simulating two-dimensional vibrational spectra. In 2018, Jung, Videla, and Batista\cite{Jung_Batista:2018} showed that the response function (\ref{eq:R_qm_pert}) is related to the DKT correlation function,\cite{Tong_Sun:2020}
\begin{equation}
K(t_2,t_1) = \frac{1}{\beta^2}\int_0^{\beta} d\lambda_1\int_{0}^{\lambda_1} d\lambda_2 \text{Tr} \left[ \rho \hat{A}(-i\hbar\lambda_1) \hat{B}(t_1-i\hbar\lambda_2) \hat{C}(t_1+t_2) \right],
\end{equation}
through the frequency-domain expression
\begin{equation}
R(\omega_2, \omega_1) = Q_{+}(\omega_2,\omega_1) K^{\text{sym}}(\omega_2,\omega_1) + Q_{-}(\omega_2,\omega_1) K^{\text{asym}}(\omega_2,\omega_1), \label{eq:full_dkt_rel}
\end{equation}
where
\begin{eqnarray}
K^{\text{sym}}(t_2,t_1) = 2 \text{Re} [K(t_2,t_1)], \qquad K^{\text{asym}}(t_2,t_1) = 2 i \text{Im} [K(t_2,t_1)],\\
f(\omega_1, \omega_2) = \int_{-\infty}^{\infty} d t_1 \int_{-\infty}^{\infty} d t_2 f(t_2, t_1) e^{-i\omega_1 t_1 - i \omega_2 t_2},
\end{eqnarray}
and $Q_{\pm}$ are some known functions of $\omega_{1}$ and $\omega_2$.\cite{Tong_Sun:2020} Although Eq.~(\ref{eq:full_dkt_rel}) is exact, an additional approximation is needed before $K(t_2, t_1)$ can be evaluated using the RPMD. Specifically, the authors employed the symmetrized DKT approximation
\begin{equation}
R(\omega_2, \omega_1) \approx Q_{+}(\omega_2,\omega_1) K^{\text{sym}}(\omega_2,\omega_1), \label{eq:sym_dkt_rel}
\end{equation}
neglecting the asymmetric contribution to the DKT correlation function, to make use of the RPMD approximation
\begin{equation}
K^{\text{sym}}(t_2,t_1) \approx \int dq \int dp A_N(q) B_N(q_{t_1}) C_N(q_{t_1+t_2}) \rho_N(q, p).
\end{equation}
Thus, the symmetrized DKT approximation, and as a consequence the RPMD DKT method as well, is not exact in the classical limit. In addition, the corresponding response functions need not be zero at $t_2=0$,\cite{Tong_Sun:2020} except in the high-temperature limit, in which case the RPMD DKT method reduces to the classical correlation function approach of DeVane, Ridley, Space, and Keyes.\cite{DeVane_Keyes:2003,DeVane_Keyes:2004,DeVane_Keyes:2005,DeVane_Keyes:2006}

\section{Results and discussion}

In this Section, the above considerations about the classical, RPMD, and RPMD DKT methods for simulating two-dimensional vibrational spectra are demonstrated numerically on an anharmonic model system determined by the Hamiltonian
\begin{equation}
H_{1\rm{D}}(q, p) = \frac{1}{2}(p^2 + q^2) + a q^3 + a^2 q^4, \label{eq:pot}
\end{equation}
where $a$ is a tunable parameter that controls the degree of anharmonicity. This system can be solved numerically exactly in a finite basis and was previously used with $a=0.1$ to evaluate the accuracy of RPMD and nonequilibrium RPMD methods.\cite{Craig_Manolopoulos:2004,Welsch_Miller:2016} Details of the exact quantum and approximate trajectory-based simulations can be found in the Supplementary Material.

\begin{figure}[pth]
\includegraphics[width=\textwidth]{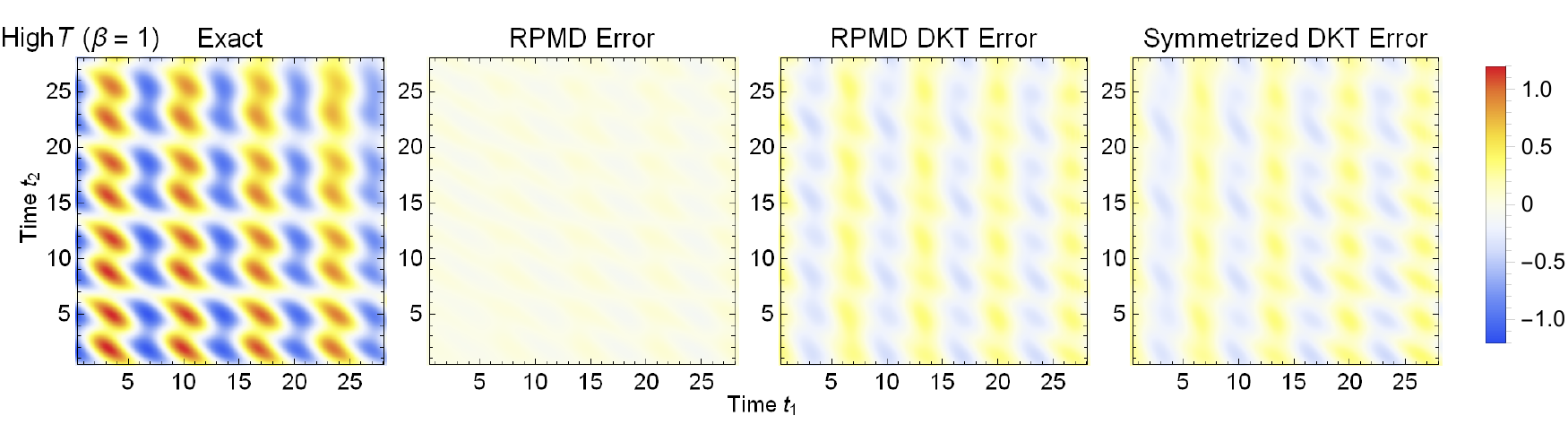}
\includegraphics[width=\textwidth]{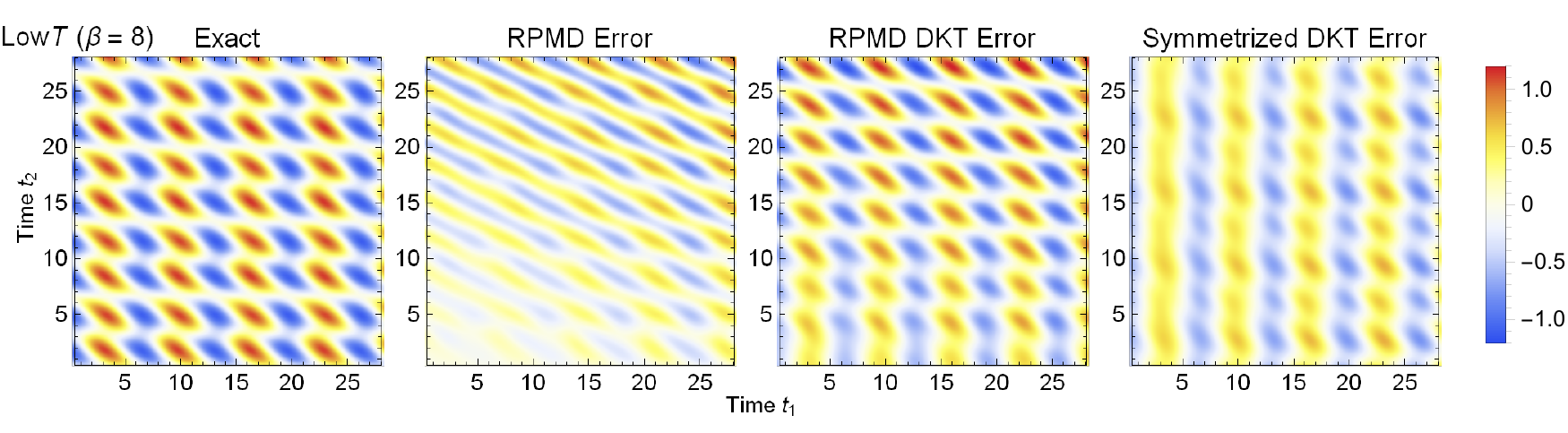}
\caption{Errors of approximate methods (columns 2--4) for evaluating the two-time response function (shown in the left-most column): New equilibrium-nonequilibrium RPMD method (labeled ``RPMD''), RPMD DKT approximation,\cite{Jung_Batista:2018} and the exact quantum symmetrized DKT correlation function (``Symmetrized DKT'').\cite{Jung_Batista:2018} Results are shown at both high ($\beta = 1$, top) and low ($\beta=8$, bottom) temperatures, with the anharmonicity parameter $a=0.1$ [Eq.~(\ref{eq:pot})], and operators $\hat{A}=\hat{B}=\hat{q}$, $\hat{C} = \hat{q}^2 / 2$.}
\label{fig:plot2D}
\end{figure}

Figure~\ref{fig:plot2D} shows that the equilibrium-nonequilibrium RPMD approach performs better than the RPMD DKT method at both low and high temperatures. Notably, the newly proposed RPMD approach exhibits the expected short-time accuracy, while most of the error in the RPMD DKT method can be attributed to the neglect of the asymmetric part of the full DKT correlation function (compare columns 3 and 4 of Fig.~\ref{fig:plot2D}).

\begin{figure}[pth]
\includegraphics[width=\textwidth]{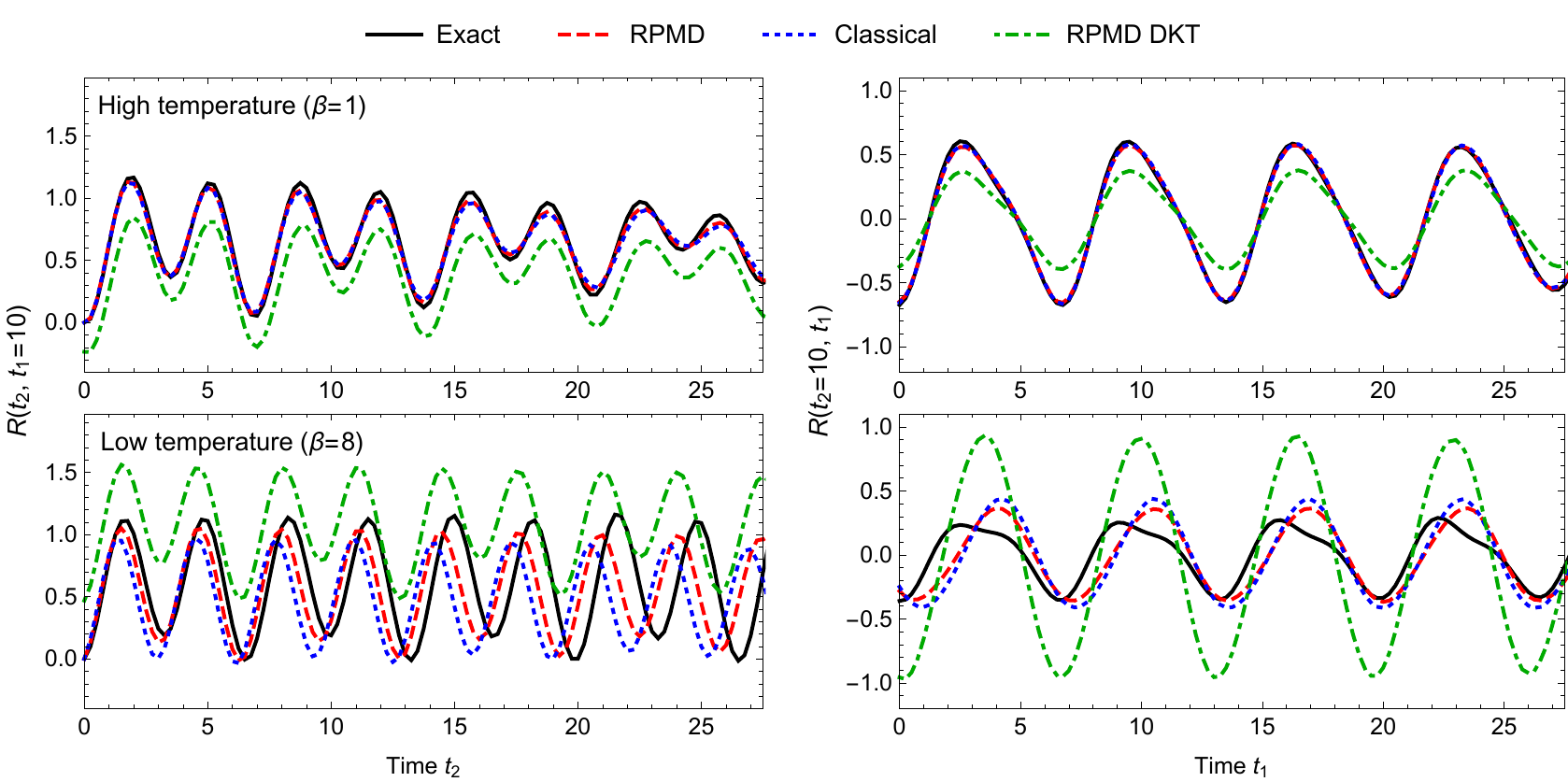}
\centering
\caption{Cuts along $t_1 = 10$ (left) and $t_2 = 10$ (right) of the two-dimensional response function at $\beta = 1$ (top) and $\beta=8$ (bottom) evaluated with the exact quantum approach, with the RPMD and classical equilibrium-nonequilibrium methods, and with the RPMD DKT approach. System's parameters and operators were same as in Fig.~\ref{fig:plot2D}.}
\label{fig:plot_cuts}
\end{figure}

Figure~\ref{fig:plot_cuts} shows the time slices of $R(t_2, t_1)$ along $t_2$ (left panels) and $t_1$ (right panels). At low temperature (Fig.~\ref{fig:plot_cuts}, bottom), equilibrium-nonequilibrium RPMD is more accurate than the classical method due to the inclusion of nuclear quantum effects, while the two are similar at high temperature (Fig.~\ref{fig:plot_cuts}, top). Although the RPMD response function agrees with the exact result at short time (see Fig.~\ref{fig:plot_cuts}, bottom left), its long-time oscillation frequency deviates from the exact, because RPMD is not expected to recover quantum coherence effects. In this example, the classical simulation is more accurate than the RPMD DKT method, showing that the uncontrolled error of the symmetrized DKT approximation can outweigh the benefits of including nuclear quantum effects. In fact, similar findings were implied, even if not explicitly discussed, by the earlier reports on the symmetrized DKT approximation.\cite{Jung_Batista:2018,Tong_Sun:2020} In addition, both classical and RPMD equilibrium-nonequilibrium methods are exact (zero) at $t_2 = 0$ (see left panels of Fig.~\ref{fig:plot_cuts}), which does not hold for the RPMD DKT approach (see also Figs.~6e and 8e of Ref.~\onlinecite{Tong_Sun:2020}).

\begin{figure}[pth]
\includegraphics[width=0.5\textwidth]{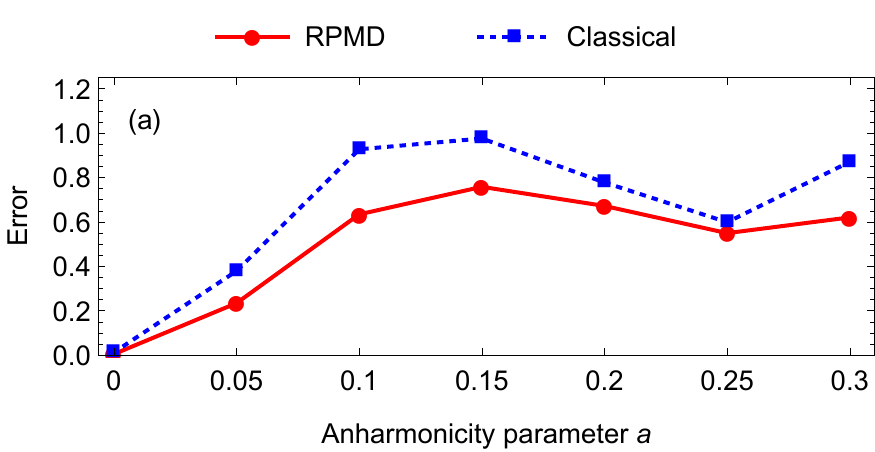}
\includegraphics[width=0.5\textwidth]{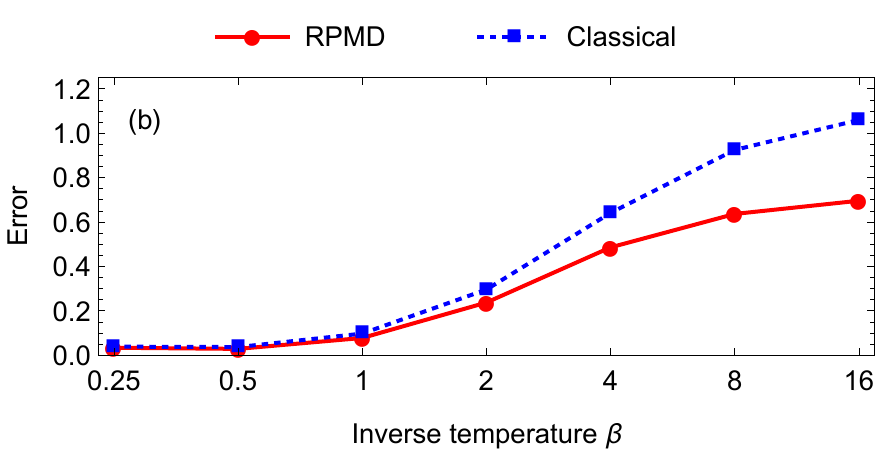}
\includegraphics[width=0.5\textwidth]{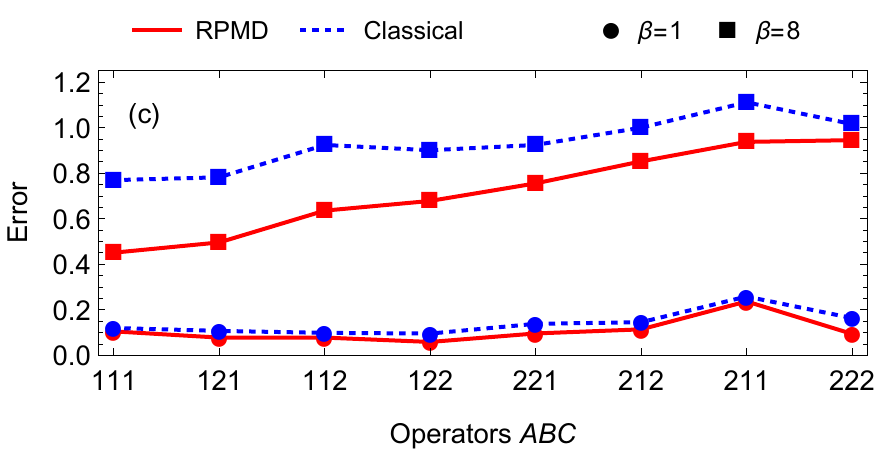}
\centering
\caption{Errors based on Eq.~(\ref{eq:error}) of the approximate two-time response functions for systems with different degree of anharmonicity (a), at different temperatures (b), and for different combinations of linear and quadratic operators $\hat{A}$, $\hat{B}$, and $\hat{C}$ (c). Apart from the parameters that are being varied or explicitly indicated in each plot, the default parameters are $a=0.1$ [Eq.~(\ref{eq:pot})], $\beta=8$, $\hat{A}=\hat{B}=\hat{q}$, $\hat{C} = \hat{q}^2 / 2$. In panel (c), ``1'' denotes linear operator $\hat{q}$, ``2'' denotes quadratic operator $\hat{q}^2 / 2$. For example, ``122'' means $\hat{A} = \hat{q}$ and $\hat{B} = \hat{C} = \hat{q}^2/2$. Operator combinations are ordered according to the error of the RPMD simulation at $\beta = 8$.}
\label{fig:anh_tem_op}
\end{figure}

To demonstrate numerically that the proposed approach inherits some well-known properties of RPMD, we choose to quantify the error of an approximate response function $R^{\text{approx}}$ as
\begin{equation}
\text{Error} = \frac{\lVert R^{\text{approx}} - R^{\text{exact}}\rVert}{\text{max}(\lVert R^{\text{approx}} \rVert, \lVert R^{\text{exact}}\rVert)}, \label{eq:error}
\end{equation}
where $\lVert R \rVert^2 = \int dt_1 \int dt_2 |R(t_2, t_1)|^2$ and $R^{\text{exact}}$ is the exact response function. Figure~\ref{fig:anh_tem_op} shows that the equilibrium-nonequilibrium RPMD consistently outperforms the classical approach for systems of different anharmonicities (a), at different temperatures (b), and for different combinations of operators (c). As discussed earlier, the RPMD method is exact for the harmonic oscillator ($a=0$, see Fig.~\ref{fig:anh_tem_op}a) and certain combinations of operators $\hat{A}$, $\hat{B}$, and $\hat{C}$. Interestingly, the classical approach is also exact in some of those limiting cases (see Supplementary Material). Nevertheless, as soon as the system is even weakly anharmonic, RPMD is clearly superior to the classical approach. As expected, both RPMD and classical methods perform worse as the anharmonicity increases. Furthermore, the equilibrium-nonequilibrium RPMD method converges to the exact (classical) result in the high-temperature limit ($\beta \rightarrow 0$, Fig.\ref{fig:anh_tem_op}b). Again, as the temperature decreases, the RPMD approach becomes less accurate. Finally, the proposed RPMD approach is more accurate for linear compared to nonlinear operators (Fig.~\ref{fig:anh_tem_op}c).

\begin{figure}[pth]
\includegraphics[width=0.8\textwidth]{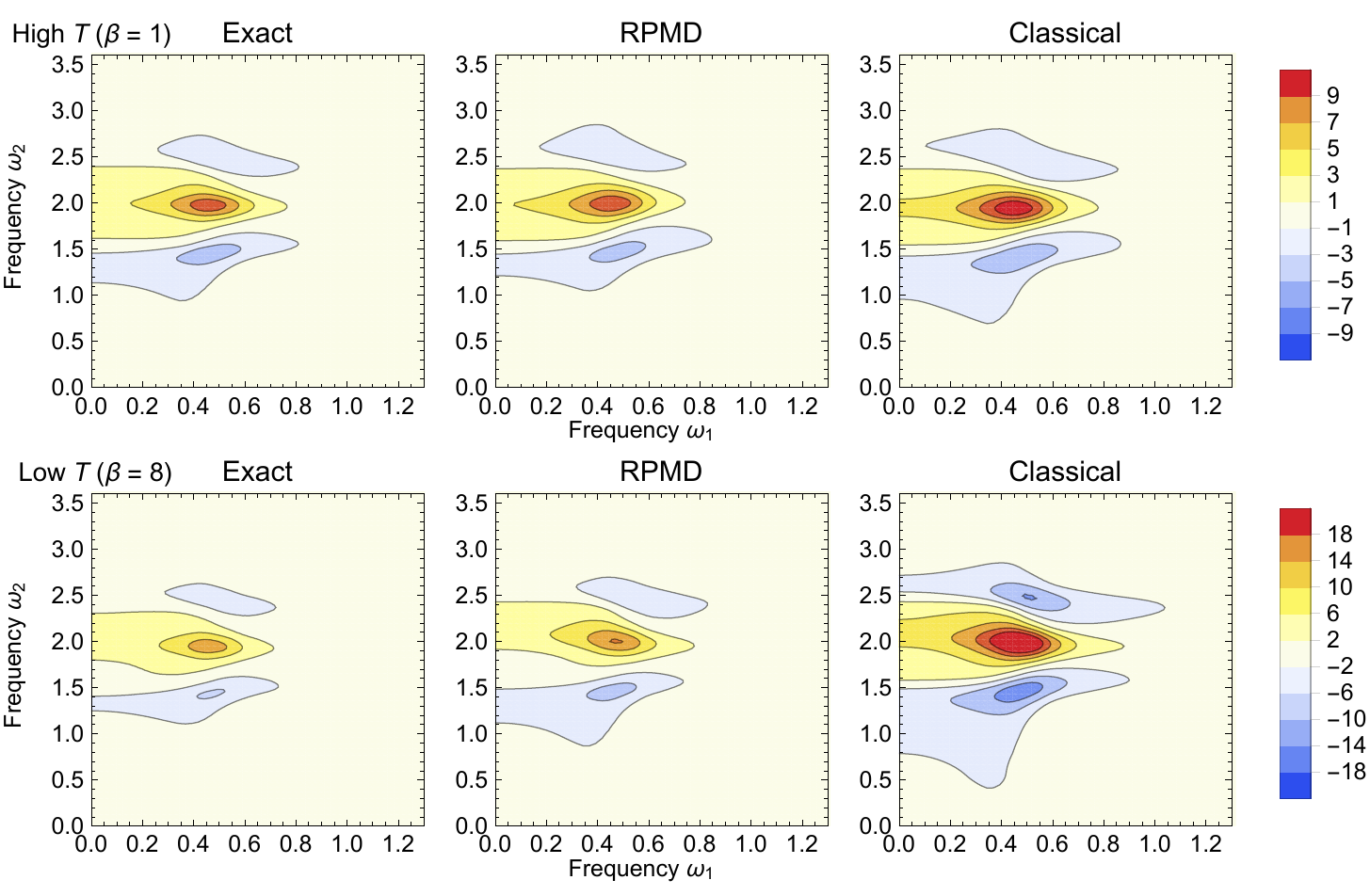}
\caption{Exact (left), equilibrium-nonequilibrium RPMD (center), and equilibrium-nonequilibrium classical MD (right) two-dimensional spectra of the two-dimensional model system described by the Hamiltonian~(\ref{eq:pot_2d}) with $\hat{A} = \hat{q}_1$ and $\hat{B}=\hat{C}=\hat{q}_2$, simulated at high ($\beta = 1$, top) and low ($\beta = 8$, bottom) temperatures.}
\label{fig:plot2D_spectra}
\end{figure}

We further consider a two-dimensional model Hamiltonian
\begin{equation}
H_{2\rm{D}}(q_1, p_1, q_2, p_2) = H_{1\rm{D}}(\Omega_1 q_1, p_1) + H_{1\rm{D}}(\Omega_2 q_2, p_2) + \lambda q_1 q_2 \label{eq:pot_2d}
\end{equation}
composed of two anharmonic oscillators [Eq.~(\ref{eq:pot}), $a=0.2$] with different central frequencies ($\Omega_1 = 0.5$, $\Omega_2 = 2$) and a linear-linear coupling term proportional to $\lambda = 0.1$. The light-matter interaction operators are set to $\hat{A}=\hat{q}_1$, $\hat{B}=\hat{C}=\hat{q}_2$, reflecting the terahertz-infrared-Raman pulse sequence,\cite{Grechko_Bonn:2018,Vietze_Grechko:2021} in which the first, terahertz pulse interacts only with the low-frequency mode $q_1$, while the infrared and off-resonant Raman interactions probe the high-frequency mode $q_2$. Figure~\ref{fig:plot2D_spectra} shows that the two-dimensional spectrum, obtained as the double cosine transform of the two-time response function (see Supplementary Material for further details), exhibits off-diagonal peaks at $(\Omega_1, \Omega_2)$ and $(\Omega_1, \Omega_2 \pm \Omega_1)$, as predicted by the Feynman diagrams of Refs.~\onlinecite{Sidler_Hamm:2020,Vietze_Grechko:2021}. The signal vanishes in the absence of coupling, $\lambda = 0$, or anharmonicity, $a=0$ (see Section~III of the Supplementary Material). Equilibrium-nonequilbrium RPMD and classical MD spectra reproduce the shape of the exact spectrum. However, at low temperature (Fig.~\ref{fig:plot2D_spectra}, bottom), classical method largely overestimates the magnitudes of the off-diagonal peaks due to the neglect of the nuclear quantum effects. In our simplified case, this results mostly in an overall scaling factor, but, in general, could affect the relative intensities of the spectral features.

\section{Conclusion}

To conclude, we have introduced an RPMD-based method that captures nuclear quantum effects on three-pulse two-dimensional vibrational spectra at the cost of classical MD simulations. The proposed approach was shown to perform better than the classical and recently developed RPMD DKT methods. Although our discussion focused only on RPMD, the overall scheme could be extended to the more general Matsubara dynamics\cite{Hele_Althorpe:2015,Hele_Althorpe:2015a,Hele:2017} and to its efficient approximations, including thermostated RPMD,\cite{Rossi_Manolopoulos:2014} CMD,\cite{Jang_Voth:1999} and quasi-CMD.\cite{Trenins_Althorpe:2019} It would be interesting to see how these approximations perform when applied to multidimensional spectroscopy, because it is expected to provide a more stringent test for dynamical approximations compared to linear spectroscopy, for which similar studies exist.\cite{Benson_Althorpe:2019} Because two-dimensional Raman and hybrid THz--Raman spectra vanish for harmonic potentials with linear dipole moment or polarizability operators,\cite{Tanimura_Mukamel:1993} adequate computational approaches must capture the effect of electrical and mechanical anharmonicities, which poses a challenge for RPMD and the above-mentioned methods.\cite{Ple_Bonella:2021,Benson_Althorpe:2021}

To derive the equilibrium-nonequilibrium RPMD, we did not invoke the concept of the Kubo transformation, which is traditionally done for one-time correlation functions,\cite{Habershon_Miller:2013} but instead combined nonequilibrium RPMD with classical response theory. This new way of deriving real-time path-integral methods will be of special interest for simulating time-resolved spectroscopic experiments, but also in other contexts where multi-time correlation or response functions appear, or if a derivation of the appropriate Kubo transformed correlation function is not obvious. The theory is general and enables new applications of established quantum dynamical methods.

\section*{Supplementary material}

See the supplementary material for the computational details, analytical expressions in the harmonic limit, and further discussion of the two-mode system in the absence of coupling or anharmonicity.

\begin{acknowledgments}
The authors thank Kenneth A. Jung, Roman Korol, and Jorge L. Rosa-Ra\'{i}ces for helpful discussions. TB acknowledges financial support from the Swiss National Science Foundation through the Early Postdoc Mobility Fellowship (grant number P2ELP2-199757). GAB and TFM gratefully acknowledge support from the National Science Foundation Chemical Structure, Dynamics and Mechanisms program (grant CHE-1665467). The computations presented here were conducted in the Resnick High Performance Computing Center, a facility supported by Resnick Sustainability Institute at the California Institute of Technology.
\end{acknowledgments}

\section*{Data availability\label{sec:data}}

The data that support the findings of this study are available from the corresponding author upon reasonable request.

\bibliographystyle{aipnum4-2}
\bibliography{bibliography}

\end{document}


\title{Supplementary material for: Equilibrium-nonequilibrium ring-polymer molecular dynamics for nonlinear spectroscopy}
\author{Tomislav Begu\v{s}i\'{c}}
\email{tbegusic@caltech.edu}
\author{Xuecheng Tao}
\affiliation{Division of Chemistry and Chemical Engineering, California Institute of Technology, Pasadena, California 91125, USA}
\author{Geoffrey A. Blake}
\affiliation{Division of Chemistry and Chemical Engineering, California Institute of Technology, Pasadena, California 91125, USA}
\affiliation{Division of  Geological and Planetary Sciences, California Institute of Technology, Pasadena, California 91125, USA}
\author{Thomas F. Miller III}
\affiliation{Division of Chemistry and Chemical Engineering, California Institute of Technology, Pasadena, California 91125, USA}
\date{\today}

\begin{abstract}

\end{abstract}

\maketitle

\section{Computational details}

\subsection{One-dimensional model}

\begin{figure}[hbp]
\includegraphics[width=0.6\textwidth]{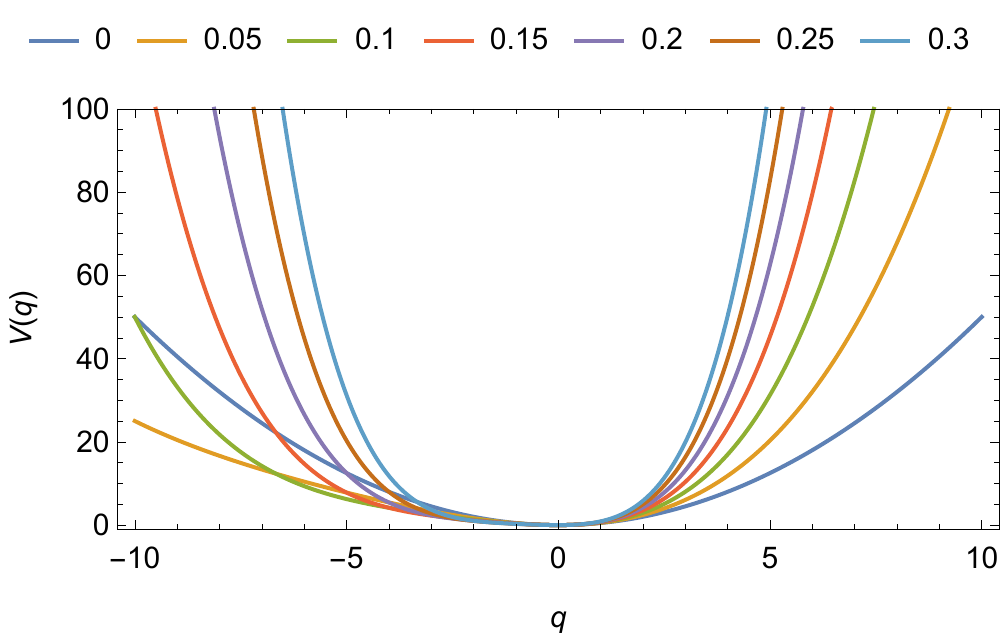}
\caption{Potentials $V(q) = 0.5 q^2 + a q^3 + a^2 q^4$ of varying degree of anharmonicity, controlled by the parameter $a$, plotted on the position grid used in quantum calculations.}
\label{fig:plot_potentials}
\end{figure}

\begin{figure}[pth]
\includegraphics[width=\textwidth]{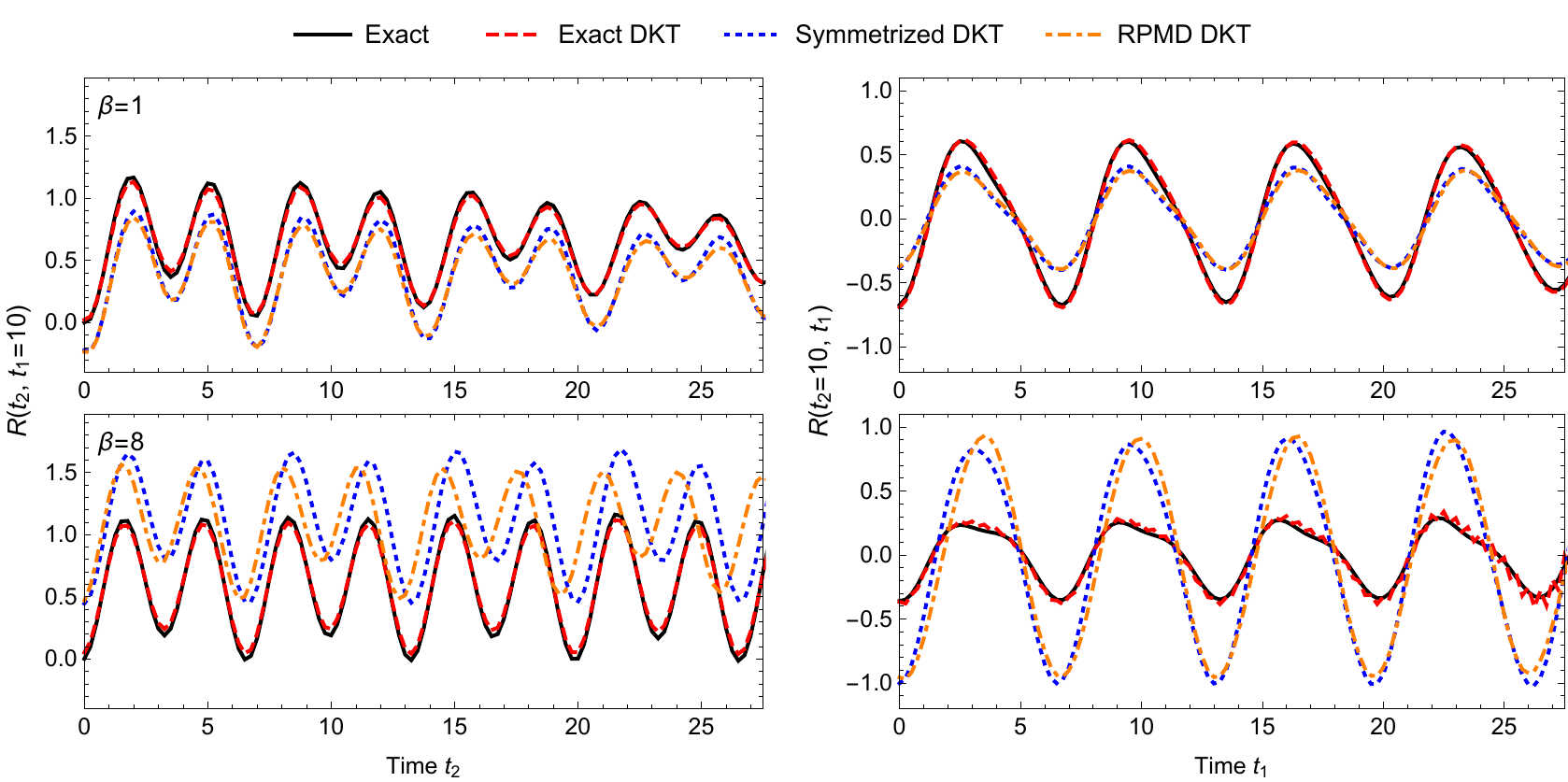}
\caption{Same as Fig.~2 of the main text, but comparing two exact quantum simulations of the response function, using either the direct evaluation or the DKT correlation function, and two approximate calculations based on the symmetrized DKT correlation function and its RPMD approximation.}
\label{fig:plot_cuts_DKT}
\end{figure}

Here, the response function $R(t_2, t_1)$ was computed for 120 steps along both time coordinates with a time step of 0.25, with the exact result obtained by solving the time-independent Schr\"{o}dinger equation on a position grid (see Fig.~\ref{fig:plot_potentials}) ranging from -10 to 10 with a step of 0.01 and transforming the operators $\hat{A}$, $\hat{B}$, and $\hat{C}$ to the eigenstate basis. Depending on the temperature, between 10 and 60 states were used in the evaluation of the response function. The same approach was used for computing the exact DKT correlation function. Since the relation between the DKT correlation function $K(t_2, t_1)$ and $R(t_2,t_1)$ assumes a Fourier transform involving both positive and negative $t_1$ and $t_2$, DKT correlation functions were computed for positive and negative times, i.e., from -30 to 30. In addition, the Fourier transform assumes infinite integral boundaries, which can be replaced by final times only if the integrand is zero beyond some time. This is usually the case because correlation functions are typically damped by the environment. In our simple system, no damping is applied to the correlation functions, which is why the relation between the computed $K(t_2, t_1)$ and $R(t_2, t_1)$ holds only approximately. To avoid artifacts due to boundary effects, the Fourier transform of the DKT correlation function, $K(\omega_2, \omega_1)$, was multiplied by $\exp[-0.02(\omega_1^2 + \omega_2^2)]$. For the same reason, only $t_1, t_2 < 27.5$ are used to offer a fair comparison between DKT-based and other methods. Figure~\ref{fig:plot_cuts_DKT} shows that the numerical error introduced in this way is negligible.

\begin{figure}
\includegraphics[width=0.6\textwidth]{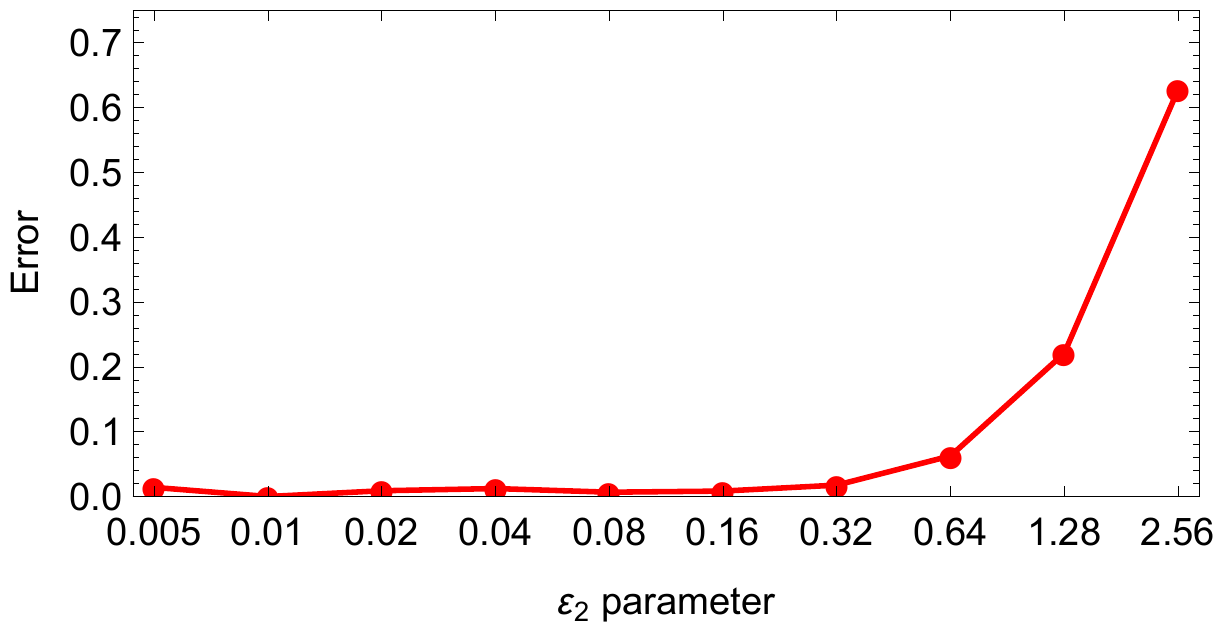}
\caption{Effect of the $\varepsilon_2$ parameter on the equilibrium-nonequilibrium RPMD response function. Error is measured by Eq.~(38) of the main text with $R^{\text{exact}}(t_2, t_1)$ replaced by the RPMD response function computed with $\varepsilon_2 = 0.01$. Figure shows that the results are not affected by the choice of $\varepsilon_2$ for a wide range of values.}
\label{fig:epsilon}
\end{figure}

To obtain classical and RPMD results, $K = 2^{17} \approx 10^5$ initial conditions $(q^{(k)},p^{(k)})$ were sampled from a thermal trajectory computed using the Andersen thermostat.\cite{Andersen:1980} Then, four trajectories were run for each initial condition: two nonequilibrium trajectories starting from $(q^{(k)}, p^{(k)} \pm (\varepsilon_2 / 2) N [\partial B_N(q^{(k)})/\partial q^{(k)}])$ with $\varepsilon = 0.01$ (see Fig.~\ref{fig:epsilon}), a forward trajectory starting from $(q^{(k)}, p^{(k)})$, and a backward trajectory that is computed by propagating $(q^{(k)},-p^{(k)})$ forward in time. Trajectories were propagated using a second-order symplectic algorithm\cite{Ceriotti_Manolopoulos:2010} with a time step of 0.05 for 600 steps, while the observables were evaluated every 5 steps of the dynamics. Classical results were obtained by setting the number of beads $N = 1$, while $N=64$ was used for the converged RPMD results. The nonequilibrium and backward trajectories were used for evaluating the equilibrium-nonequilibrium response function according to
\begin{equation}
R(t_2, t_1) = - \frac{\partial}{\partial t_1} \left\lbrace \frac{\beta}{K \varepsilon_2}\sum_{k=1}^{K} [C_N(q_{+, t_2}^{(k)}) - C_N(q_{-, t_2}^{(k)})] A_N(q_{-t_1}^{(k)}) \right\rbrace,
\end{equation}
where the time derivative was computed numerically. The forward and backward equilibrium trajectories were used for evaluating the RPMD DKT response function for positive and negative $t_1$ and $t_2$ times.

\subsection{Two-dimensional model}

The two-time response function of the two-dimensional model was evaluated for 200 steps (time step 0.25) in $t_1$ and $t_2$. Exact results were obtained in the same way as for the one-dimensional system, using a two-dimensional grid between $-20$ and $20$ (step $0.01$) along each coordinate and the first $50$ eigenstates. For the RPMD ($N=64$) and classical ($N=1$) simulations, we averaged the response function over $2^{17}$ samples and the trajectories were run for 1000 steps (time step 0.05, observables evaluated every 5 steps). External field parameter was $\varepsilon_2 = 0.01$.

\begin{figure}[pth]
\includegraphics[width=\textwidth]{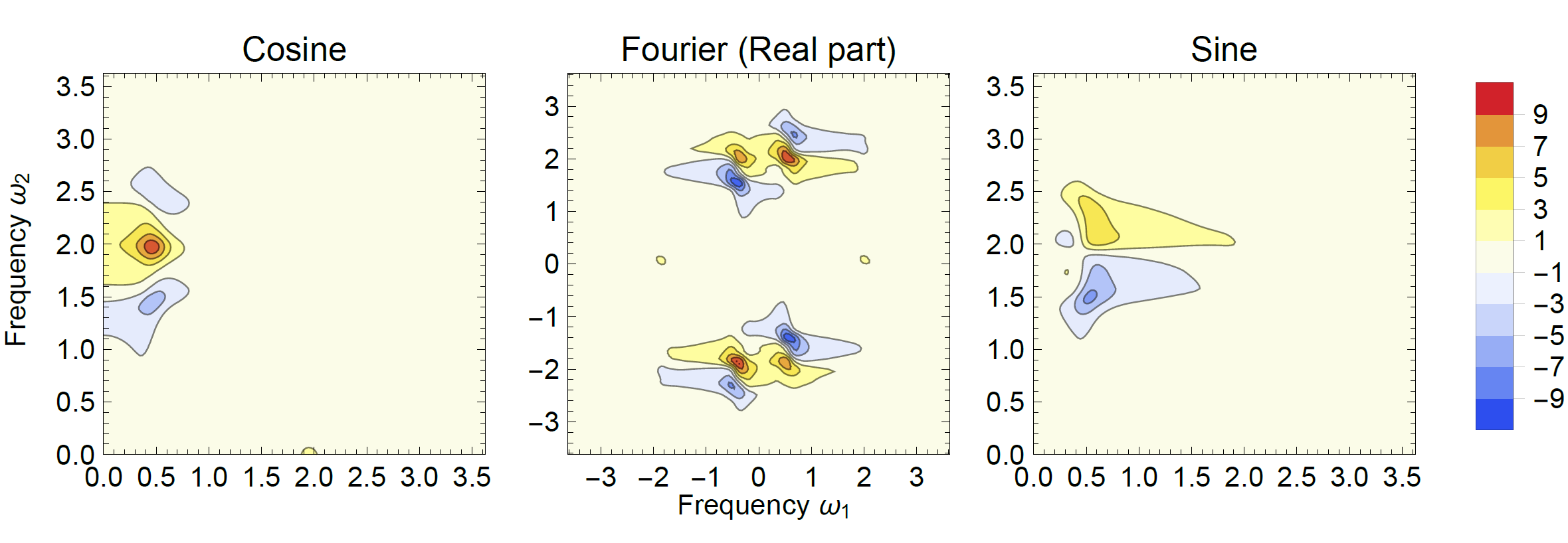}
\caption{Cosine [Eq.~(\ref{eq:S_cos})], Fourier [Eq.~(\ref{eq:S_F}), real part], and sine [Eq.~(\ref{eq:S_sin})] transforms of the damped two-time response function of the two-mode system, presented in the main text, evaluated with exact quantum dynamics at $\beta=8$.}
\label{fig:plot2D_transforms}
\end{figure}

Spectra for the two-dimensional model were evaluated as
\begin{equation}
S_C(\omega_2, \omega_1) = \int_{0}^{\infty} dt_2 \int_{0}^{\infty} dt_1 R(t_2, t_1) e^{-(t_1 + t_2)/\tau} \cos(\omega_1 t_1) \cos(\omega_2 t_2), \label{eq:S_cos}
\end{equation}
where $\tau=7.5$. Before taking the discrete cosine transform, the damped response function was padded with zeros up to $t_1,t_2 = 1300$. The double cosine transform of Eq.~(\ref{eq:S_cos}) is related to the Fourier transform
\begin{equation}
S_F(\omega_2, \omega_1) = \int_{0}^{\infty} dt_2 \int_{0}^{\infty} dt_1 R(t_2, t_1) e^{-(t_1 + t_2)/\tau} e^{-i \omega_1 t_1 - i\omega_2 t_2} \label{eq:S_F}
\end{equation}
by
\begin{equation}
S(\omega_2, \omega_1) = \frac{1}{2}\text{Re}[ S_F(\omega_2, \omega_1) + S_F(-\omega_2, \omega_1)].\label{eq:S_F_sym}
\end{equation}
An alternative is to use the sine transform
\begin{align}
S_S(\omega_2, \omega_1) 
&= \int_{0}^{\infty} dt_2 \int_{0}^{\infty} dt_1 R(t_2, t_1) e^{-(t_1 + t_2)/\tau} \sin(\omega_1 t_1) \sin(\omega_2 t_2) \label{eq:S_sin}\\
&= -\frac{1}{2}\text{Re}[ S_F(\omega_2, \omega_1) - S_F(-\omega_2, \omega_1)],
\end{align}
which was proposed in Ref.~\onlinecite{Hamm_Savolainen:2012} as a way to extract the purely absorptive component of the two-dimensional spectrum. We show these different transforms in Fig.~\ref{fig:plot2D_transforms}. In this case, the main off-diagonal peak at $(\omega_1, \omega_2) \approx (\Omega_1=0.5,\Omega_2=2.0)$ does not appear in the sine transform, but is well resolved in the cosine transform. The peaks appearing in the Fourier transform spectrum (Fig.~\ref{fig:plot2D_transforms}, middle) can be assigned to the Feynman diagrams of Ref.~\onlinecite{Vietze_Grechko:2021}.

\section{Harmonic limit}

In this Section, we show that the RPMD method proposed in the main text is exact in the limit $N \rightarrow \infty$ for the harmonic oscillator,
\begin{equation}
V(q) = \frac{1}{2} m \omega^2 q^2,
\end{equation}
and when two of the operators $\hat{A}$, $\hat{B}$, and $\hat{C}$ are linear in position.

\subsection{Quantum response functions}

To derive the exact quantum results, we first rewrite
\begin{eqnarray}
R(t_2, t_1) 
&=& -\frac{1}{\hbar^2} \text{Tr}\left\lbrace \hat{C}(t_2 + t_1) [\hat{B}(t_1), [\hat{A}, \hat{\rho}]] \right\rbrace \label{eq:R_qm_pert_1} \\
&=& -\frac{1}{\hbar^2} \text{Tr}\left\lbrace [ [ \hat{C}(t_2 + t_1), \hat{B}(t_1)], \hat{A}] \hat{\rho} \right\rbrace \label{eq:R_qm_pert_2} \\
&=& -\frac{1}{\hbar^2} \left \langle [ [ \hat{C}(t_2 + t_1), \hat{B}(t_1)], \hat{A}] \right\rangle \label{eq:R_qm_pert_3}
\end{eqnarray}
and note that
\begin{equation}
[\hat{q}(t), f(\hat{q})] = \frac{1}{m \omega} [\hat{p}, f(\hat{q})] = -\frac{i\hbar}{m\omega}f^{\prime}(\hat{q}) \sin \omega t, \label{eq:comm}
\end{equation}
where $f^{\prime}(q) = \partial f / \partial q$.

Then, for $\hat{A} = \hat{B} = \hat{q}$ and $\hat{C} = C(\hat{q})$, we have
\begin{eqnarray}
R(t_2, t_1)
&=& -\frac{1}{\hbar^2} \left \langle [ [ C(\hat{q}), \hat{q}(-t_2)], \hat{q}(-t_1 - t_2)]\right\rangle \\
&=& -\frac{1}{\hbar^2} \left( -\frac{i\hbar}{m \omega} \right) \sin \omega t_2 \left \langle [ C^{\prime}(\hat{q}), \hat{q}(-t_1 - t_2)]\right\rangle \\
&=& \frac{i}{m \hbar \omega} \sin \omega t_2 \left( -\frac{i\hbar}{m \omega} \right) \sin \omega (t_1 + t_2) \left \langle C^{\prime \prime}(\hat{q})\right\rangle \\
&=& \frac{\langle C^{\prime \prime}(\hat{q})\rangle}{(m \omega)^2} \sin \omega t_2 \sin \omega (t_1 + t_2), \label{eq:R_AB_Q}
\end{eqnarray}
where we applied Eq.~(\ref{eq:comm}) twice. For $\hat{A} = \hat{C} = \hat{q}$ and $\hat{B} = B(\hat{q})$:
\begin{eqnarray}
R(t_2, t_1)
&=& -\frac{1}{\hbar^2} \left \langle [ [ \hat{q}(t_2), B(\hat{q})], \hat{q}(-t_1)]\right\rangle\\
&=& -\frac{1}{\hbar^2} \left( -\frac{i\hbar}{m \omega} \right) \sin \omega t_2 \left \langle [ B^{\prime}(\hat{q}), \hat{q}(-t_1)]\right\rangle \\
&=& \frac{\langle B^{\prime \prime}(\hat{q})\rangle}{(m \omega)^2} \sin \omega t_1 \sin \omega t_2. \label{eq:R_AC_Q}
\end{eqnarray}
For $\hat{B} = \hat{C} = \hat{q}$ and $\hat{A} = A(\hat{q})$:
\begin{eqnarray}
R(t_2, t_1)
&=& -\frac{1}{\hbar^2} \left \langle [ [ \hat{q}(t_2), \hat{q}], \hat{A}(-t_1)]\right\rangle\\
&=& -\frac{1}{\hbar^2} \left \langle \left [ \left( - \frac{i\hbar}{m\omega} \right) \sin \omega t_2, \hat{A}(-t_1) \right]\right\rangle\\
&=& 0, \label{eq:R_BC_Q}
\end{eqnarray}
because scalars commute with operators.

\subsection{RPMD response functions}

Assuming $\varepsilon_2$ is small, we can derive the fully equilibrium RPMD from the equilibrium-nonequilibrium RPMD:
\begin{eqnarray}
R(t_2, t_1) 
&=& \frac{\beta}{\varepsilon_2} \langle [ C_N(q_{+, t_2}) - C_N(q_{-, t_2})] \dot{A}_N(q_{-t_1})\rangle, \\
&=& N \beta \left \langle \left[\frac{ \partial C_N(q_{t_2})}{\partial q_{t_2}}\right]^{T} \cdot M_{qp, t_2} \cdot \frac{\partial B_N(q)}{\partial q} \dot{A}_N(q_{-t_1}) \right \rangle, \label{eq:RPMD_eq}
\end{eqnarray}
where we used
\begin{eqnarray}
C_N(q_{\pm, t_2}) 
&\approx& C_N(q_{t_2}) \pm \frac{\varepsilon_2}{2}\left[ \frac{\partial C_N(q_{t_2})}{\partial p} \right]^{T} \cdot N \frac{\partial B_N(q)}{\partial q} \\
&\approx& C_N(q_{t_2}) \pm \frac{\varepsilon_2}{2} N \left[\frac{ \partial C_N(q_{t_2})}{\partial q_{t_2}}\right]^{T} \cdot \frac{\partial q_{t_2}}{\partial p} \cdot \frac{\partial B_N(q)}{\partial q}
\end{eqnarray}
and defined $M_{xy, t} = \partial x_t / \partial y$. Equation~(\ref{eq:RPMD_eq}) provides yet another way to employ RPMD in the simulation of two-time response function $R(t_2, t_1)$. Similar to the hybrid equilibrium-nonequilibrium approach, the equilibrium method also reduces to its classical MD analogue.\cite{Saito_Ohmine:2002,Saito_Ohmine:2003,Hasegawa_Tanimura:2006} However, this method is less practical due to the need for computing and converging the stability matrix element $M_{qp, t_2}$, which has been deemed difficult even in the classical MD setting. Finally, we note that the equilibrium-nonequilibrium and fully equilibrium methods give equal results when $\varepsilon_2$ is small.

Before continuing with specific cases, we rewrite Eq.~(\ref{eq:RPMD_eq})
\begin{equation}
R(t_2, t_1) = N \beta \left\langle \left[\frac{ \partial \tilde{C}_N(\tilde{q}_{t_2})}{\partial \tilde{q}_{t_2}}\right]^{T} \cdot M_{\tilde{q}\tilde{p}, t_2} \cdot \frac{\partial \tilde{B}_N(\tilde{q})}{\partial \tilde{q}} \dot{\tilde{A}}_N(\tilde{q}_{-t_1}) \right\rangle, \label{eq:RPMD_eq_norm_modes}
\end{equation}
in terms of the ring-polymer normal mode coordinates $\tilde{q} = O \cdot q$, where $O$ is an orthogonal $N \times N$ matrix that diagonalizes the ring-polymer force-constant and $\tilde{f}(\tilde{q}) = f(q)$. Specifically, the first normal mode $\tilde{q}_1 = \sqrt{N} q_c$ is a scalar multiple of the ring-polymer centroid $q_c = \frac{1}{N}\sum_{i=1}^{N} q_i$ and evolves according to
\begin{equation}
\tilde{q}_{1, t} = \tilde{q}_1 \cos \omega t + \frac{1}{m \omega} \tilde{p}_1 \sin \omega t.
\end{equation}

Then, for $A(q) = B(q) = q$ and arbitrary $C(q)$, we have:
\begin{eqnarray}
R(t_2, t_1) 
&=& N \beta \left \langle \frac{\partial \tilde{C}_N(\tilde{q}_{t_2})}{\partial \tilde{q}_{1, t_2}} M_{\tilde{q}_1\tilde{p}_1, t_2} \frac{1}{\sqrt{N}} \dot{\tilde{A}}_N(\tilde{q}_{-t_1}) \right \rangle \label{eq:R_AB_1} \\
&=& \frac{\sqrt{N} \beta}{m \omega} \sin \omega t_2 \left \langle \frac{\partial \tilde{C}_N(\tilde{q}_{t_2})}{\partial \tilde{q}_{1, t_2}} \dot{\tilde{A}}_N(\tilde{q}_{-t_1}) \right \rangle \label{eq:R_AB_2} \\
&=& \frac{\sqrt{N} \beta}{m \omega} \sin \omega t_2 \left \langle \frac{\partial \tilde{C}_N(\tilde{q})}{\partial \tilde{q}_{1}} \dot{\tilde{A}}_N(\tilde{q}_{-t_1 - t_2}) \right \rangle \label{eq:R_AB_3} \\
&=& \frac{\beta}{m} \sin \omega t_2 \sin \omega (t_1 + t_2) \left \langle \frac{\partial \tilde{C}_N(\tilde{q})}{\partial \tilde{q}_{1}} \tilde{q}_1 \right \rangle \label{eq:R_AB_4} \\
&=& \frac{N}{(m\omega)^2} \left \langle \frac{\partial^2 \tilde{C}_N(\tilde{q})}{\partial \tilde{q}_{1}^2} \right \rangle  \sin \omega t_2 \sin \omega (t_1 + t_2) \label{eq:R_AB_5} \\
&=& \frac{1}{(m\omega)^2} \left \langle \frac{1}{N} \sum_{i=1}^{N}\frac{\partial^2 C(q_i)}{\partial q_{i}^2} \right \rangle  \sin \omega t_2 \sin \omega (t_1 + t_2),  \label{eq:R_AB_6} \\
&=& \frac{\langle C^{\prime \prime} \rangle^{\text{RP}}}{(m\omega)^2} \sin \omega t_2 \sin \omega (t_1 + t_2),  \label{eq:R_AB_7}
\end{eqnarray}
where we used $\tilde{A}_N(\tilde{q}) = \tilde{B}_N(\tilde{q}) = q_c$, $\partial q_c / \partial \tilde{q}_1 = 1 / \sqrt{N}$, and $\partial \tilde{q}_{i, t} / \partial \tilde{p}_1 = \delta_{i1}\sin \omega t / (m \omega)$ in Eqs.~(\ref{eq:R_AB_1}) and (\ref{eq:R_AB_2}), $\dot{\tilde{A}}_N(\tilde{q}_t) = - (\omega / \sqrt{N}) \tilde{q}_1 \sin \omega t + (1/m\sqrt{N})\tilde{p}_1 \cos \omega t$ to derive Eq.~(\ref{eq:R_AB_4}), in which we immediately dropped the second term that vanishes, because $\langle p_1 \rangle = 0$,  and $\tilde{q}_1 \exp(-\beta_N H_N) = -(N/\beta m \omega^2) (\partial / \partial \tilde{q}_1)\exp(-\beta_N H_N)$ together with the integration by parts to obtain Eq.~(\ref{eq:R_AB_5}). Finally, Eq.~(\ref{eq:R_AB_6}) follows from Eq.~(\ref{eq:R_AB_5}) because
\begin{equation}
\frac{\partial^2 \tilde{C}_N(\tilde{q})}{\partial \tilde{q}_{1}^2} = \left(\frac{\partial q}{\partial \tilde{q}_1}\right)^T \cdot \frac{\partial^2 C_N(q)}{\partial q^2} \cdot \frac{\partial q}{\partial \tilde{q}_1} = \frac{1}{N} e^T \cdot \frac{\partial^2 C_N(q)}{\partial q^2} \cdot e = \frac{1}{N^2} \sum_{i=1}^{N} \frac{\partial^2 C(q_i)}{\partial q_i},
\end{equation}
where we used $\partial q / \partial \tilde{q}_1 = e / \sqrt{N}$ and defined the $N$-dimensional vector $e = (1 \dots 1)^T$. In the limit of $N \rightarrow \infty$, Eq.~(\ref{eq:R_AB_6}) converges to the quantum result (\ref{eq:R_AB_Q}). Similarly, for $A(q) = C(q) =q$ and arbitrary $B(q)$:
\begin{eqnarray}
R(t_2, t_1) 
&=& N \beta \left \langle \frac{1}{\sqrt{N}} \frac{1}{m\omega} \sin \omega t_2 \frac{\partial \tilde{B}_N(\tilde{q})}{\partial \tilde{q}_1} \frac{\omega}{\sqrt{N}} \tilde{q}_1 \sin \omega t_1 \right \rangle \label{eq:R_AC_1} \\
&=& \frac{\beta}{m} \left \langle \frac{\partial \tilde{B}_N(\tilde{q})}{\partial \tilde{q}_1} \tilde{q}_1 \right \rangle \sin \omega t_1 \sin \omega t_2 \label{eq:R_AC_2} \\
&=& \frac{ \langle B^{\prime \prime} \rangle^{\text{RP}} }{(m\omega)^2} \sin \omega t_1 \sin \omega t_2, \label{eq:R_AC_3} 
\end{eqnarray}
which reduces to the corresponding quantum expression (\ref{eq:R_AC_Q}) when converged with respect to the number of beads $N$. Finally, for $B(q) = C(q) = q$,
\begin{eqnarray}
R(t_2, t_1)
&=& N \beta \left\langle \frac{1}{\sqrt{N}} \frac{1}{m\omega} \sin \omega t_2 \frac{1}{\sqrt{N}} \dot{\tilde{A}}_N(\tilde{q}_{-t_1}) \right\rangle \\
&=& \frac{1}{m \omega} \sin \omega t_2 \left\langle  \dot{\tilde{A}}_N(\tilde{q}_{-t_1}) \right\rangle = 0,
\end{eqnarray}
because equilibrium expectation values are time-independent.

Interestingly, in certain cases, even the classical method ($N=1$) is exact. This is trivially true for the last studied case [$B(q) = C(q) = q$], but also for certain non-trivial cases, e.g., for $C(q) \propto q^2$ in the first studied example [$A(q) = B(q) = q $] and $B(q) \propto q^2$ in the second studied case [$A(q) = C(q) = q$] because the corresponding final expressions are independent of temperature. This can be seen in Fig.~3a of the main text for the anharmonicity parameter $a=0$, $A(q) = B(q) = q$, and $C(q) = q^2/2$, where both classical and RPMD results are exact.

\section{Two-mode system in the absence of coupling or anharmonicity}

In the main text we showed spectra for the system described by the Hamiltonian $\hat{H}_{2\rm{D}}$ [see Eq.~(37) of the main text] and operators $\hat{A} = \hat{q}_1$ and $\hat{B} = \hat{C} = \hat{q}_2$. The corresponding two-time response function
\begin{equation}
R(t_2, t_1) 
= -\frac{1}{\hbar^2} \text{Tr}\left\lbrace \hat{q}_{2}(t_2 + t_1) [\hat{q}_{2}(t_1), [\hat{q}_1, \hat{\rho}]] \right\rbrace \label{eq:R_2D_full}
\end{equation}
can be shown to vanish in the absence of coupling ($\lambda = 0$) or anharmonicity ($a=0$). In the limit $\lambda = 0$, $\hat{\rho} = \hat{\rho}_1 \hat{\rho}_2$ and, therefore,
\begin{equation}
R(t_2, t_1) 
= -\frac{1}{\hbar^2} \text{Tr}_{q_2}\left\lbrace \hat{q}_{2}(t_2 + t_1) [\hat{q}_{2}(t_1), \hat{\rho}_2] \right\rbrace \text{Tr}_{q_1}\left\lbrace [\hat{q}_1, \hat{\rho}_1] \right\rbrace
=0,
\end{equation}
because the trace of a commutator is zero. If $a=0$, the system is harmonic, which means it can be decomposed into normal modes. Since all operators are linear in $q_1$ and $q_2$, they will also be linear in the normal-mode coordinates and, as shown in the previous Section, the response function is zero for linear operators in a harmonic potential.

\bibliographystyle{aipnum4-2}
\bibliography{bibliography}